\documentclass{article}
\usepackage{amssymb}
\usepackage{amsmath}
\usepackage{cite}
\usepackage[notref,notcite]{showkeys}
\usepackage{pstricks}
\allowdisplaybreaks
\numberwithin{equation}{section}
\DeclareMathOperator{\sgn}{\rm sgn}
\DeclareMathOperator{\diag}{\rm diag}

\DeclareMathOperator{\reg}{\rm reg}
\DeclareMathOperator{\Rs}{\mathbb{R}}

\DeclareMathOperator{\Cs}{\mathbb{C}}
\DeclareMathOperator{\Go}{\mathcal{G}}
\DeclareMathOperator{\Lo}{\mathcal{L}}
\DeclareMathOperator{\No}{\mathcal{N}}
\DeclareMathOperator{\Do}{\mathcal{D}}
\DeclareMathOperator{\Vo}{\mathcal{V}}
\DeclareMathOperator{\Po}{\mathcal{P}}

\DeclareMathOperator{\bk}{\mathbf{k}}
\def\t#1{\widetilde{#1}}
\def\h#1{\widehat{#1}}
\newtheorem{theorem}{Theorem}[section]
\newtheorem{lemma}[theorem]{Lemma}

\newtheorem{remark}{Remark}[section]

\begin{document}

\title{Extended resolvent of heat operator with multisoliton potential}
\author{M.~Boiti${}^{*}$, F.~Pempinelli${}^{*}$, and A.~K.~Pogrebkov$
{}^{\dag}$ \\
${}^{*}$Dipartimento di Matematica e Fisica, Universit\`a del Salento and\\
Sezione INFN, Lecce, Italy\\
${}^{\dag}$Steklov Mathematical Institute, Moscow, Russia}
\date{MSC: 37K10, 37K15, 35C08, 37K40
Extended Resolvent, Solitons}
\maketitle

\begin{abstract}
The heat operator with a general multisoliton potential is considered and its extended resolvent, depending on a parameter $q\in\Rs^2$ is derived. Its boundedness properties in all variables and its discontinuities in the parameter $q$ are given. As the result, the Green's functions are introduced and their properties are studied in detail.
\end{abstract}

\section{Introduction}

The Kadomtsev--Petviashvili equation in its version called KPII
\begin{equation}
(u_{t}-6uu_{x_{1}}+u_{x_{1}x_{1}x_{1}})_{x_{1}}=-3u_{x_{2}x_{2}},
\label{KPII}
\end{equation}
where $u=u(x,t)$, $x=(x_{1},x_{2})$ and subscripts $x_{1}$, $x_{2}$ and $t$ denote partial derivatives, is a (2+1)-dimensional generalization of the celebrated Korteweg--de~Vries (KdV) equation. The KPII equations, originally derived as a model for small-amplitude, long-wavelength, weakly two-dimensional waves in a weakly dispersive medium~\cite{KP1970}, have been
known to be integrable since the beginning of the 1970s~\cite{D1974,ZS1974}, and can be considered as a prototypical (2+1)-dimensional integrable equation.

The KPII equation is integrable via its association with the operator
\begin{equation}
\mathcal{L}(x,\partial_{x})=-\partial_{x_{2}}+\partial_{x_{1}}^{2}-u(x), \label{heatop}
\end{equation}
which define the well known equation of heat conduction, or heat equation for short. The spectral theory of the operator (\ref{heatop}) was developed in \cite{BarYacoov,Lipovsky,Wickerhauser,Grinevich0} in the case of a real potential $u(x)$ rapidly decaying at spatial infinity, which, however, is not the most interesting case, since the KPII equation was just proposed in \cite{KP1970} in order to deal with two dimensional weak transverse perturbation of the one soliton solution of the KdV. The main difficulty in the study of this problem is due to the fact that the soliton solutions of the KPII equations are not decaying at space infinity and have a ray behavior on the $x$-plane (see, e.g., \cite{BPPPr2001a,asympKPII,ChK2,K}). Correspondingly, the integral equations defining the Jost solutions in this case are senseless, as their kernels do not exist. A spectral theory of the KPII equation that also includes solitons has to be build, as already successfully done for the KPI equation \cite{BPP2006b}. Following a framework of this article we develop a generalization of the standard IST so called ``scattering on nontrivial background,'' i.e., we consider a potential
\begin{equation}
\t{u}(x)=u(x)+ u'(x),\qquad x=(x_1,x_2),\label{tu}
\end{equation}
where $u(x)$ is some multisoliton potential and $u'(x)$ is a smooth, rapidly enough decaying function of its variables, that can be considered a perturbation of the soliton potential. Correspondingly, the Jost solution of the operator $\t{\Lo}$ with potential $\t{u}$ can be defined as the solution of the following integral equation
\begin{equation}
\t\Phi(x,\bk)=\Phi(x,\bk)+\int dy\,\Go(x,y,\bk)u'(y)\t\Phi(y,\bk),\label{tphi}
\end{equation}
where $\bk\in\Cs$ is a spectral parameter, $\Phi(x,\bk)$ is the Jost solution of the operator $\Lo$ with multisoliton potential $u$ and $\Go(x,y,\bk)$ is its total Green's function, i.e.,
\begin{equation}
\bigl(-\partial_{x_{2}}^{}+\partial_{x_{1}}^{2}-u(x)\bigr)\Go(x,x',\bk)=\delta (x-x').  \label{green}
\end{equation}
In order  to be useful for developing the IST on the solitonic background, the Green's function must obey condition of boundedness, i.e., function
\begin{equation}
G(x,x',\bk)=e_{}^{i\bk(x_{1}-x_{1}')+\bk^{2}(x_{2}-x_{2}')}\Go(x,x',\bk)  \label{Green}
\end{equation}
must bounded with respect to the variables $x,x'\in \Rs^{2}$ and $\bk\in\Cs$ and has finite limits at infinity. In this framework it was possible to develop the inverse scattering transform for a solution describing a perturbation of the one soliton solution~\cite{BPPP2002}. However, the case of any number of solitons is still open. For this case, in \cite{KPIIGreen} we derived a total Green's function, which is a natural generalization of the Green's function for the case of a decaying potential. But as it was shown in~\cite{BPPP2002} and~\cite{BPP2006b}, in order to control the singularities of the Jost solutions we need some other Green's functions. Thus, following~\cite{BPP2006b}, in order to deal with a heat operator with a generic multisoliton potential (for some multisoliton solutions this was done in~\cite{BPPP2009}), we introduce its extended resolvent, which is an object more generic than the Green's function. Exactly, we introduce a two dimensional real spectral parameter $q=(q_{1,}q_{2})$ and consider the \textbf{extended} Lax operator
\begin{equation}
\mathcal{L}(x,\partial_{x}+q)=-\partial_{x_{2}}-q_{2}+(\partial_{x_{1}}+q_{1})^{2}-u(x),  \label{heatop2}
\end{equation}
and, then, the extended resolvent of the heat operator (\ref{heatop}) is defined as the tempered distribution $M(x,x';q)$ with respect to all its six variables $x,x',q$ that satisfies the differential equations
\begin{equation}
\mathcal{L}(x,\partial_{x}+q)M x,x';q) =\mathcal{L}^{\text{d}}(x',\partial_{x'}+q)M(x,x';q) =\delta (x-x'),  \label{risolv1}
\end{equation}
where $\mathcal{L}^{\text{d}}(x,\partial_{x})$ is the dual of $\mathcal{L}(x,\partial_{x})$.

The extended resolvent can be considered the generating functional of the different Green's functions of the $\Lo$-operator (\ref{heatop}). Indeed, let us introduce
\begin{equation}
\widehat{M}( x,x';q) =e^{q(x-x')}M(x,x';q),  \label{Mhat}
\end{equation}
that, of course, is not necessarily a tempered distribution. Nevertheless, it is easy to see that
\begin{equation}
\mathcal{L}(x,\partial_{x})\widehat{M}(x,x';q)=\mathcal{L}^{\text{d}}(x',\partial_{x'})\widehat{M}(x,x';q)=
\delta(x-x').  \label{MhatGreen}
\end{equation}
In particular, the total Green's function is given by means of the reduction
\begin{equation}
\Go(x,x',\bk)=\h{M}(x,x';q)\Biggm|_{\substack{q_{1}=\bk_{\Im}\\ q_{2}=\bk_{\Im}^2-\bk_{\Re}^2}}.\label{GM}
\end{equation}

The construction of the distribution $M(x,x';q)$ is the subject of this article. In \cite{asympJostKPII} it was shown that the heat operator (\ref{heatop}) with a multisoliton potential $u(x)$ can have left or right annihilators in polygonal regions of the $q$-plane, where, consequently, the resolvent cannot exists. Here we give an explicit expression for the extended resolvent and prove that outside some special  polygon region (the same as was given in~\cite{asympJostKPII}) it exists as a tempered distribution and obeys  (\ref{risolv1}). We demonstrate that in this region reduction (\ref{GM}) is always possible for an arbitrary $\bk\in\Cs$ and it gives the total Green's function derived in~\cite{KPIIGreen}. Thus the condition that $M(x,x';q)$ is a distribution leads to the boundedness property (\ref{Green}). According to the procedure followed in the case of the nonstationary Schr\"{o}dinger operator in \cite{BPP2006b}, the description of the discontinuities of the total Green's function on the complex plane of the spectral parameter requires some special reductions of the extended resolvent, i.e., auxiliary Green's functions. Thus we conclude this article by a detailed study of these discontinuities.

\section{Heat operator with multisoliton potential and its Jost solutions}

Soliton potentials (see \cite{BPPP2009} and \cite{asympJostKPII,BPPPr2001a,asympKPII,ChK2,K,equivKPII} for details) are
labeled by two numbers (topological charges) $N_{a}$ and $N_{b}$, which obey condition
\begin{equation}
N_{a},N_{b}\geq 1.  \label{nanb}
\end{equation}
Let
\begin{equation}
\No=N_{a}+N_{b},  \label{Nnanb}
\end{equation}
so that $\No\geq 2$. We introduce the $\No$ real parameters
\begin{equation}
\kappa_{1}<\kappa_{2}<\ldots <\kappa_{\No},  \label{kappas}
\end{equation}
and the functions
\begin{equation}
K_{n}^{}(x)=\kappa_{n}^{}x_{1}^{}+\kappa_{n}^{2}x_{2}^{},\quad n=1,\ldots ,\No.  \label{Kn}
\end{equation}
Let
\begin{equation}
e^{K(x)}=\diag\{e^{K_{n}(x)}\}_{n=1}^{\No}  \label{eK}
\end{equation}
be a diagonal $\No\times {\No}$ matrix, let $\Do$ be a $\No\times {N_{b}}$ constant matrix and $\Vo$ be an ``incomplete Vandermonde matrix,'' i.e., the $N_{b}\times \No$ matrix
\begin{equation}
\Vo=\left(
\begin{array}{lll}
1 & \ldots & 1 \\
\kappa_{1} & \ldots & \kappa_{\No} \\
\vdots &  & \vdots \\
\kappa_{1}^{N_{b}-1} & \ldots & \kappa_{\No}^{N_{b}-1}%
\end{array}\right) .  \label{W}
\end{equation}
Then, the soliton potential is given by
\begin{equation}
u(x)=-2\partial_{x_{1}}^{2}\log \tau (x),  \label{ux}
\end{equation}
where the $\tau$-function can be expressed as
\begin{equation}
\tau (x)=\det \bigl(\Vo e^{K(x)}\Do\bigr).  \label{tau}
\end{equation}
For the Jost and dual Jost solutions (solutions, respectively, of the heat operator (\ref{heatop}) and its dual) we have
\begin{align}
& \Phi (x,\bk)=e^{-i\bk x_{1}-\bk^{2}x_{2}}\chi (x,\bk),  \label{Phi} \\
& \Psi (x,\bk)=e^{i\bk x_{1}+\bk^{2}x_{2}}\xi (x,\bk),  \label{Psi}
\end{align}
where
\begin{align}
& \chi (x,\bk)=\dfrac{\tau_{\Phi}^{}(x,\bk)}{\tau (x)},  \label{symPhi}\\
& \xi (x,\bk)=\dfrac{\tau_{\Psi}^{}(x,\bk)}{\tau (x)},  \label{symPsi}
\end{align}
with (Miwa shift)
\begin{equation}
\tau_{\Phi}(x,\bk)=\det \bigl(\Vo e^{K(x)}(\kappa +i\bk)\Do\bigr),\qquad
\tau_{\Psi}(x,\bk)=\det \left( \Vo\dfrac{e^{K(x)}}{\kappa +i\bk}\Do\right),\label{tauk:1}
\end{equation}
and
\begin{equation}
\kappa +i\bk=\diag\{\kappa_{n}+i\bk\}_{n=1}^{\No}.  \label{kappadiag}
\end{equation}

In order to study the properties of the potential and the Jost solutions, it is convenient to use the representations for the $\tau$-functions, which follow by the Binet--Cauchy formula for the determinant of a product of matrices, i.e.,
\begin{align}
& \tau (x)=\dfrac{1}{N_{b}!}\sum_{\{n_{i}\}=1}^{\No}\Do(\{n_{i}\})V(\{n_{i}\})
\prod_{l=1}^{N_{b}}e^{K_{n_{l}}(x)},  \label{tauf1}\\
&\chi (x,\bk)=\dfrac{1}{N_{b}!\tau (x)}\sum_{\{m_{i}\}=1}^{\No}\Do(\{m_{i}\})V(\{m_{i}\})         \prod_{l=1}^{N_{b}}(\kappa_{m_{l}}+i\bk)e_{}^{K_{m_{l}}(x)},  \label{Phik1} \\
&\xi (x,\bk)=\dfrac{1}{N_{b}!\tau (x)}\sum_{\{n_{i}\}=1}^{\No}\Do(\{n_{i}\})V(\{n_{i}\}) \prod_{l=1}^{N_{b}}\dfrac{e_{}^{K_{n_{l}}(x)}}{\kappa_{n_{l}}+i\bk},  \label{Psik1}
\end{align}
where we used notations
\begin{align}
& V(\{n_{i}\})=\det \left(\begin{array}{lll}
1 & \ldots & 1 \\
\kappa_{n_{1}} & \ldots & \kappa_{n_{N_b}} \\
\vdots &  & \vdots \\
\kappa_{n_{1}}^{N_{b}-1} & \ldots & \kappa_{n_{N_b}}^{N_{b}-1}
\end{array}\right)\equiv\prod_{1\leq i<j\leq N_{b}}(\kappa_{n_{j}}-\kappa_{n_{i}}),\label{V} \\
& \Do(\{n_{i}\})=\det \left(\begin{array}{ccc}
\Do_{n_{1},1} & \dots & \Do_{n_{1},N_{b}} \\
\vdots &  & \vdots \\
\Do_{n_{N_{b}},1} & \dots & \Do_{n_{N_{b}},N_{b}}\end{array}\right)  \label{Do1}
\end{align}
for the maximal minors of matrices $\Vo$ and $\Do$ and where
\begin{equation}
\{m_{i}\}=\{m_{1},\ldots,m_{N_{b}}\},\qquad\{n_{i}\}=\{n_{1},\ldots,n_{N_{b}}\}  \label{not}
\end{equation}
stand for non ordered sets of $N_b$ indices from the interval $1,\ldots,\No$. We recall that the maximal minors of a matrix satisfy the Pl\"{u}cker relation, i.e., for any subsets $\{m_{i}\}$ and $\{n_{i}\}$ of indices running from $1$ to $\mathcal{N}$ and arbitrary $j\in \{1,\ldots ,N_{b}\}$
\begin{align}
& \Do(\{m_{i}\})\Do(\{n_{i}\})=  \notag \\
&=\sum_{s=1}^{N_{b}}\Do(m_{1},\ldots,m_{s-1},n_{j},m_{s+1},\ldots,m_{N_{b}})
\Do(n_{1},\ldots,n_{j-1},m_{s},n_{j+1},\ldots ,n_{N_{b}}).\label{plu}
\end{align}

Notice that the only $x$-dependent terms in (\ref{tauf1}), (\ref{Phik1}), and (\ref{Psik1}) are exponents of sums of linear functions (\ref{Kn}). Correspondingly, the asymptotic behavior of the function $\tau(x)$ and of the potential has a sectorial structure on the $x$-plane. In order to specify these sectors at $x\to\infty$ we introduce the ray directions
\begin{equation}  \label{rn}
r_{n}:\qquad\left\{\begin{array}{l}
x_1+(\kappa_{n}+\kappa_{n+N_b})x_2\quad\text{bounded} \\
(\kappa_{n+N_b}-\kappa_{n})x_2\to-\infty.\end{array}\right.,\quad n=1,\ldots,\No,
\end{equation}
where we assume that the indices are defined mod$\,\No$, so that thanks to (\ref{Nnanb}), say, $n+N_b=n-N_a$ for $n>N_a$. Thus there are $N_a$ rays in the direction $x_2\to-\infty$ and $N_b$ rays in the direction $x_2\to+\infty$. The sector $\sigma_{n}$ is swept out by rotating anticlockwise the ray $r_n $ up to the ray $r_{n+1}$. These sectors are nonintersecting and cover
the whole $x$-plane with the exception of rays. In \cite{asympKPII} we proved that the the leading exponents of $\tau(x)$ when $x\to\infty$ are the exponents $\exp\bigl(\sum_{l=n}^{n+N_b-1}K_{l}(x)\bigr)$, each being the leading one in the corresponding $\sigma_{n}$ sector of the $x$-plane. More exactly, if the coefficients
\begin{equation}
z_{n}=V(n,\ldots,n+N_{b}-1)\Do(n,\ldots,n+N_{b}-1),  \label{zn}
\end{equation}
are different from zero for all $n=1,\ldots,\No$ (again with indices defined mod$\,\No$) the function $\tau(x)$ has, along rays and inside sectors, the
following asymptotic behavior
\begin{align}
&x\overset{r_{n}}{\longrightarrow}\infty: & & \tau (x)=\bigl(z_{n}+z_{n+1}e_{}^{K_{N_{b}+n}(x)-K_{n}(x)}+o(1)\bigr)\exp\Biggl(
\sum_{l=n}^{n+N_{b}-1}K_{l}(x) \Biggr),  \label{3:232} \\
&x\overset{\sigma_{n}}{\longrightarrow}\infty: & & \tau(x)=\bigl(z_{n}+o(1)\bigr)
\exp\Biggl(\sum_{l=n}^{n+N_{b}-1}K_{l}(x)\Biggr).  \label{tauasympt}
\end{align}
Regularity of the potential $u(x)$ on the $x$-plane is equivalent to the absence of zeroes of $\tau(x)$. It is clear that it is enough to impose the condition that the matrix $\Do$ is Totally Non Negative (TNN), i.e., that
\begin{equation}
\Do(n_1,\ldots,n_{N_b})\geq0,\quad\text{for all}\quad 1\leq n_{1}<\ldots<n_{N_{b}}\leq\No.  \label{D}
\end{equation}
However, sufficient conditions on the matrix $\Do$ for the regularity of the potential are unknown. On the other side, from (\ref{3:232}) and (\ref{tauasympt}) it follows directly that it is sufficient to require that
\begin{equation}
z_n>0  \label{zn0}
\end{equation}
for having nonsingular asymptotics of the potential.

We also mention that the functions $\chi (x,\bk)$ and $\xi (x,\bk)$ have bounded asymptotics on the $x$-planes because the $x$-dependent exponents enter in denominators and numerators of expressions (\ref{Phik1}) and (\ref{Psik1}) with coefficients proportional to $\Do(\{n_{i}\})$. This means that the leading asymptotic behavior of the denominators of the functions $\chi(x,\bk)$ and $\xi (x,\bk)$ on the $x$-plane is not weaker than the behavior of their numerators. For more details see~\cite{BPPPr2001a,asympJostKPII,equivKPII,asympKPII}, where the same notations have been used.

We need in the following also the values $\chi (x,i\kappa_{n})$ of $\chi(x,\bk)$ at $\bk=i\kappa_{n}$ and the residues $\xi_{n}(x)$ of $\xi (x,\bk)$ at $\bk=i\kappa_{n}$. From (\ref{Phi}), (\ref{Psi}) and (\ref{Phik1}), (\ref{Psik1}) we have
\begin{align}
&\chi(x,i\kappa_{n})=\dfrac{(-1)^{N_b}}{N_{b}!\tau(x)}\sum_{\{m_{i}\}=1}^{\No} \Do(\{m_{i}\})V(\{m_{i}\},n)\prod_{l=1}^{N_{b}}e^{K_{m_{l}}(x)},  \label{1}\\
&\xi_{n}(x)=\dfrac{1}{iN_{b}!\tau(x)}\sum_{\{n_{i}\}=1}^{\No}\Do(\{n_{i}\})\sum_{j=1}^{N_{b}}
\delta_{n_{j}n}(-1)^{j-1}V(n_1,\ldots,\widehat{n_j},\ldots,n_{N_b}) \prod_{l=1}^{N_{b}}e^{K_{n_{l}}(x)},  \label{2}
\end{align}
where $\{\{m_{i}\},n\}=\{m_{1},\ldots,m_{N_{b}},n\}$, hat over $n_{j}$ denotes that this index is omitted and where the $\delta_{n_{j}n}$ Kronecker symbol in the r.h.s.\ of the last formula is due to the fact that the residues of the terms in the sum are nonzero only when some $n_{j}=n$.

Taking into account the analyticity properties of $\chi (x,\bk)$ and $\xi (x,\bk)$ in (\ref{Phik1}), (\ref{Psik1}) their product can be written in terms of the values $\chi (x,i\kappa_{n})$ and $\xi_{n}(x)$ as follows
\begin{equation}
\chi (x,\bk)\xi (x',\bk)=1+\sum_{n=1}^{\mathcal{N}}\dfrac{\chi(x,i\kappa_{n})\xi_{n}(x')}{\bk-i\kappa_{n}},  \label{chixi}
\end{equation}
which also will be useful in the following. In \cite{asympJostKPII} we demonstrated that the Jost solutions obey the Hirota bilinear identity
\begin{equation}
\sum_{n=1}^{\mathcal{N}}\Phi (x,i\kappa_{n})\Psi_{n}(x')=0,\label{sumPhiPsi}
\end{equation}
where in analogy to (\ref{2}) $\Psi_{n}(x)$ denotes the residue of $\Psi(x,\bk)$.

\section{Extended resolvent $M(x,x';q)$}

Here we prove that the resolvent $M(x,x';q)$, i.e., a tempered distribution of its 6 real variables that obeys (\ref{risolv1}), can be written as a sum of a continuous (in some sense) and a discrete part
\begin{equation}
M(x,x';q)=M_{\text{c}}(x,x';q)+M_{\text{d}}(x,x';q),\label{M}
\end{equation}
whose definitions and properties are given below. We also specify the infinite region on the $q$-plane where the property that $M(x,x';q)$ is a tempered distribution holds, and we prove, in fact, that  $M(x,x';q)$ is bounded for $q$ in this region. For both terms we use the hat-kernels notation introduced in (\ref{Mhat})
\begin{equation}
M_{\text{c}}(x,x';q)=e^{-q(x-x')}\widehat{M}_{\text{c}}(x,x';q), \qquad M_{\text{d}}(x,x';q)=
e^{-q(x-x')}\widehat{M}_{\text{d}}(x,x';q).  \label{McMDelta}
\end{equation}

So, first, we define
\begin{align}
\widehat{M}_{\text{c}}(x,x';q)& =-\dfrac{\sgn(x_{2}-x_{2}')}{2\pi}\int d\alpha \ \theta \bigl((q_{2}+
\alpha^{2}-q_{1}^{2})(x_{2}-x_{2}')\bigr)\times  \notag \\
& \times \Phi (x,\alpha +iq_{1})\Psi (x',\alpha +iq_{1}),\label{20}
\end{align}
where the Jost and dual Jost solutions $\Phi(x,\bk)$ and $\Psi(x,\bk)$ are defined in (\ref{Phi})--(\ref{symPsi}), $\int d\alpha$ denotes integration along the real axis, and $\theta$ is the step function of its argument. Properties of $M_{\text{c}}$ are given by the following lemma.

\begin{lemma}
The integral in the r.h.s.\ of (\ref{20}) exists and the function $M_{\text{c}}$ given in (\ref{McMDelta}) is a bounded function of its arguments for all $x,x',q\in\Rs^{2}$ and has finite limits at infinity.
\end{lemma}
\textsl{Proof.\/} Thanks to (\ref{Phi}), (\ref{Psi}), and (\ref{McMDelta}) we can write that
\begin{align}
M_{\text{c}}(x,x';q)&=M_{0}(x,x';q)-\nonumber\\
&-\sum_{n=1}^{\No}\chi(x,i\kappa_{n})\xi_{n}(x')
\dfrac{\sgn(x_{2}-x_{2}')}{2\pi}\int d\alpha\dfrac{\theta \bigl((q_{2}+\alpha^{2}-q_{1}^{2})(x^{}_{2}-x_{2}')\bigr)}
{\alpha+i(q_1-\kappa_{n})}\times\nonumber\\
&\times e_{}^{-i\alpha(x^{}_1-x'_1+2q_1(x^{}_2-x'_2))-(q_{2}+\alpha^{2}-q_{1}^{2})(x_{2}-x_{2}')},
\label{20:1}
\end{align}
where we used (\ref{chixi}) for the product of $\chi(x,\bk)\xi(x',\bk)$ and where $M_{0}(x,x';q)$ is the extended resolvent of the operator (\ref{heatop2}) in the case of the zero potential
\begin{align}
M_{0}(x,x';q)=-\dfrac{\sgn(x_{2}-x_{2}')}{2\pi}&\int d\alpha\,\theta\bigl((q_{2}+\alpha^{2}-q_{1}^{2})(x^{}_{2}-x_{2}')\bigr)\times\nonumber\\
&\times e_{}^{-i\alpha(x^{}_1-x'_1+2q_1(x^{}_2-x'_2))-(q_{2}+\alpha^{2}-q_{1}^{2})(x_{2}-x_{2}')}.\label{20:2}
\end{align}
Then the statement of the lemma follows directly thanks to the exponentially decreasing factors in (\ref{20:1}) and (\ref{20:2}). $\blacksquare$

Applying the heat operator (\ref{heatop}) to $\widehat{M}(x,x',q)_{\text{c}}$ in (\ref{20}) we get
\begin{equation*}
\mathcal{L}(x,\partial_{x})\widehat{M}_{\text{c}}(x,x';q)=\dfrac{\delta(x_{2}-x_{2}')}{2\pi}\int \!\!ds\,\Phi (x,s+iq_{1})
\Psi(x',s+iq_{1}).
\end{equation*}
The integral in the r.h.s.\ can be explicitly computed thanks to (\ref{Phi}) and (\ref{Psi}) and after inserting (\ref{chixi}) in it,  we get
\begin{align}
\dfrac{\delta (x_{2}-x_{2}')}{2\pi}& \int \!\!ds\,\Phi(x,s+iq_{1})\Psi (x',s+iq_{1})=\delta (x-x')-\delta(x_2^{}-x'_2)p(x,x',q_1)+  \notag \\
& +i\delta (x_{2}-x_{2}')\theta (x'_{1}-x_{1}^{})\sum_{n=1}^{\No}\Phi (x,i\kappa_{n})\Psi_{n}(x'),\label{g32}
\end{align}
where the last term annihilates thanks to (\ref{sumPhiPsi}) and where we denoted
\begin{equation}
p(x,x',q_1)=i\sum_{n=1}^{\No}\theta (q_{1}-\kappa_{n})\Phi (x,i\kappa_{n})\Psi_{n}(x').\label{Phat}
\end{equation}
Notice that, $p(x,x',q_1)$ does not belong to the space of  Schwartz distributions, as it can have exponential growth in some directions on the $x$-plane, \cite{asympJostKPII}. Thanks to (\ref{sumPhiPsi}) it  can also be rewritten in any of the following forms
\begin{equation}
p(x,x',q_1)=\dfrac{i}{2}\sum_{n=1}^{\No}\sgn (q_{1}-\kappa_{n})\Phi (x,i\kappa_{n})\Psi_{n}(x')
\equiv-i\sum_{n=1}^{\No}\theta(\kappa_{n}-q_{1})\Phi (x,i\kappa_{n})\Psi_{n}(x'),\label{symPhat}
\end{equation}
and
\begin{equation}
p(x,x',q_1)=0,\quad \text{for all}\quad q_1\notin[\kappa_{1},\kappa_{\mathcal{N}}].\label{5.3}
\end{equation}
Notice also that, just by definition,
\begin{equation}
\mathcal{L}(x,\partial_{x})p(x,x',q_1)=0\quad\text{for any}\quad q\in\Rs^2.\label{5.31}
\end{equation}
Thus we define the second term in (\ref{M}) as (see also (\ref{McMDelta})
\begin{equation}
\widehat{M}_{\text{d}}(x,x';q) =\mp\theta \bigl(\pm(x_{2}-x_{2}')\bigr)p(x,x',q_1),  \label{MDelta}
\end{equation}
so that for any choice of sign in the r.h.s.\ we have thanks to (\ref{5.31})
\begin{equation*}
\mathcal{L}(x,\partial_{x})\widehat{M}_{\text{d}}(x,x';q)=\delta(x_2^{}-x'_2)p(x,x'q_1),
\end{equation*}
that together with (\ref{g32}) proves that
\begin{equation}
\widehat{M}(x,x';q)=\widehat{M}_{\text{c}}(x,x';q)+\widehat{M}_{\text{d}}(x,x';q),  \label{M+MDelta}
\end{equation}
obeys (\ref{MhatGreen}), or $M(x,x';q)$ given in (\ref{M}) obeys (\ref{risolv1}) thanks to (\ref{Mhat}). Proof of the second equalities in (\ref{risolv1}) and (\ref{MhatGreen}) is analogous. Let us notice that thanks to (\ref{5.3})
\begin{equation}
M_{\text{d}}(x,x';q)=0,\quad \text{for all}\quad q_1\notin[\kappa_{1},\kappa_{\mathcal{N}}].\label{5.32}
\end{equation}

Thus in order to prove that $M(x,x';q)$ is the extended resolvent, we need to prove that ${M}_{\text{d}}(x,x';q)$ belongs to the class of tempered distributions and to specify the choice of signs in (\ref{MDelta}). For this aim we need to make explicit the dependence of ${M}_{\text{d}}(x,x';q)$ on its variables. First, we consider function $p(x,x',q_1)$ defined in (\ref{Phat}). Inserting in the r.h.s.\ (\ref{1}) and (\ref{2}), thanks to antisymmetry of minors of matrices $\Do$ and $\Vo$ (see (\ref{V}) and (\ref{Do1})) and after summing over $n$, we get
\begin{align}
p(x,x',q_1)&=\dfrac{-1}{N_{b}!(N_{b}-1)!\tau(x)\tau (x')}
\sum_{\{m_{i}\}=1}^{\No}\sum_{\{n_{i}\}=1}^{\No}\Do(\{m_{i}\})\Do(\{n_{i}\})
\theta (q^{}_{1}-\kappa_{n_{N_{b}}^{}})\times  \notag \\
& \qquad \times V(\{m_{i}\},n_{N_{b}}^{})V(n_{1},\ldots,n_{N_{b}-1}^{})\times  \notag \\
& \qquad \times \exp \Biggl(\sum_{l=1}^{N_{b}}K_{m_{l}}(x)+K_{n_{N_{b}}^{}}(x)+
\sum_{l=1}^{N_{b}-1}K_{n_{l}}(x')\Biggr).  \label{4}
\end{align}
Next, we substitute the r.h.s.\ of (\ref{plu}) with $j=N_{b}$ for the product of minors of matrix $\Do$ and exchange $m_{s}\leftrightarrow{n_{N_{b}}}$ for $s=1,\ldots,N_{b}$. Notice that under this transformation the first Vandermonde determinant changes sign, while the second Vandermonde determinant is unchanged, as well as the exponent. Thus, we have
\begin{align}
p(x,x',q_1)&=\dfrac{1}{N_{b}!(N_{b}-1)!\tau(x)\tau(x')}\sum_{s=1}^{N_{b}}\sum_{\{m_{i}\}=1}^{\No}
\sum_{\{n_{i}\}=1}^{\No}\Do(\{m_{i}\})\Do(\{n_{i}\})\times  \notag \\
& \quad \times \theta(q^{}_{1}-\kappa_{m_{s}})V(\{m_{i}\},n_{N_{b}})V(n_{1},\ldots ,n_{N_{b}-1})
\times  \notag \\
& \quad\times\exp\Biggl(\sum_{l=1}^{N_{b}}K_{m_{l}}(x)+K_{n_{N_{b}}^{}}(x)+
\sum_{l=1}^{N_{b}-1}K_{n_{l}}(x')\Biggr).  \label{5}
\end{align}
Exchanging now $m_{s}\leftrightarrow {m_{N_{b}}^{}}$ we get, summing over $s$, $N_{b}$ equal terms. Finally, we multiply (\ref{4}) by $N_{b}$, sum up with (\ref{5}) and we divide this sum by $N_{b}+1$ getting
\begin{align}
p(x,x',q_1)&= \dfrac{1}{((N_{b}-1)!)^{2}(N_{b}+1)\tau(x)\tau(x')}\sum_{\{m_{i}\}=1}^{\No}\sum_{\{n_{i}\}=1}^{\No}
\Do(\{m_{i}\})\Do(\{n_{i}\})\times  \notag \\
&\times[\theta(q_{1}-\kappa_{m_{N_{b}}})-\theta(q_{1}-\kappa_{n_{N_{b}}})]V(\{m_{i}\},n_{N_{b}})
V(n_{1},n_{2},\ldots,n_{N_{b}-1})\times  \notag \\
&\quad\times\exp\Biggl(\,{\sum_{l=1}^{N_{b}}K_{m_{l}}(x)+K_{n_{N_{b}}}(x)+
\sum_{l=1}^{N_{b}-1}K_{n_{l}}(x')}\,{\Biggr)},  \label{5.1}
\end{align}
so that for $\kappa_{n}\leq{q_1}\leq{\kappa_{n+1}}$, $n=1,\ldots,\No-1$
\begin{align}
p(x,x',q_1)&= \dfrac{1}{((N_{b}-1)!)^{2}(N_{b}+1)\tau(x)\tau(x')}\sum_{\substack{m_{i},n_{i}=1,\ldots,\No\\
[\kappa_{m_{N_b}},\kappa_{n_{N_b}}]\supseteq[\kappa_{n},\kappa_{n+1}]}}
\Do(\{m_{i}\})\Do(\{n_{i}\})\times  \notag \\
&\times[\theta(q_{1}-\kappa_{m_{N_{b}}})-\theta(q_{1}-\kappa_{n_{N_{b}}})]V(\{m_{i}\},n_{N_{b}})
V(n_{1},n_{2},\ldots,n_{N_{b}-1})\times  \notag \\
&\quad\times\exp\Biggl(\,{\sum_{l=1}^{N_{b}}K_{m_{l}}(x)+K_{n_{N_{b}}}(x)+
\sum_{l=1}^{N_{b}-1}K_{n_{l}}(x')}\,{\Biggr)},  \label{5.2}
\end{align}
where the summation runs on all $m_{1},\ldots,m_{N_b}$ $n_{1},\ldots,n_{N_b}$ from 1 to $\No$ such that the interval $[\kappa_{n},\kappa_{n+1}]$ belongs to the interval $[\kappa_{m_{N_b}},\kappa_{n_{N_b}}]$. Notice that thanks to $V(\{m_{i}\},n_{N_{b}})$ this sum does not contains terms with $m_{N_b}=n_{N_b}$.

Thanks to (\ref{McMDelta}) and (\ref{MDelta}) we have that for $\kappa_{n}\leq{q_1}\leq{\kappa_{n+1}}$, $n=1,\ldots,\No-1$
\begin{align}
M_{\text{d}}(x,x';q)&=\dfrac{\mp\theta\bigl(\pm(x_{2}-x_{2}')\bigr)e^{-q(x-x')}}{((N_{b}-1)!)^{2}(N_{b}+1)\tau(x)\tau(x')}
\sum_{\substack{m_{i},n_{i}=1,\ldots,\No\\ [\kappa_{m_{N_b}},\kappa_{n_{N_b}}]\supseteq[\kappa_{n},\kappa_{n+1}]}}
\Do(\{m_{i}\})\Do(\{n_{i}\})\times  \notag \\
&\times[\theta(q_{1}-\kappa_{m_{N_{b}}})-\theta(q_{1}-\kappa_{n_{N_{b}}})]
V(\{m_{i}\},n_{N_{b}})V(n_{1},n_{2},\ldots,n_{N_{b}-1})\times  \notag \\
&\times\exp\Biggl(\,{\sum_{l=1}^{N_{b}}K_{m_{l}}(x)+K_{n_{N_{b}}}(x)+\sum_{l=1}^{N_{b}-1}K_{n_{l}}(x')}\,{\Biggr)}
\label{6}
\end{align}
with the same condition on summation. This representation for $M_{\text{d}}(x,x';q)$ gives another proof of (\ref{5.32}).

In \cite{asympJostKPII} we demonstrated that the extended operator (\ref{heatop2}) can have annihilators when $q$ belongs to some polygons on the $q$-plane. Thus, the inverse of this operator (the extended resolvent) cannot exist for any value of $q$. We introduce on the $q$-plane the polygon $\mathcal{P}$ inscripted in the parabola $q_{2}=q_{1}^{2}$ of the $q$-plane (see Fig.~\ref{figure}), with vertices at the points $(\kappa_{n},\kappa_{n}^{2})$, for $n=1,\dots ,\mathcal{N}$, whose characteristic function is given by
\begin{equation}
\epsilon (q)=\sum_{m=1}^{\mathcal{N}-1}[\theta (q_{1}-\kappa_{n+1})-\theta(q_{1}-\kappa_{n})]
[\theta (q_{n,n+1})-\theta (q_{1\mathcal{N}})],\label{19}
\end{equation}
where
\begin{equation}
q_{mn}=q_{2}-(\kappa_{m}+\kappa_{n})q_{1}+\kappa_{m}\kappa_{n}.\label{qmn}
\end{equation}
It is obvious that this polygon divides the strip $\kappa_1<q_1<\kappa_{\No}$ on the $q$-plane in two disconnected parts. Moreover, this polygon consists of substripes given by subsequent values of $\kappa$'s as follows
\begin{equation}
\kappa_{n}<q_1<\kappa_{n+1},\qquad q_{n,n+1}>0, \quad n=1,\ldots,\No-1,\qquad q_{1,\No}<0.\label{sub}
\end{equation}
\begin{figure}[ht]\label{figure} \begin{center}
\begin{pspicture}[linewidth=1.0pt](-4,-1)(4,7)
\parabola[linestyle=dashed](-4,6)(0,0)\rput[lB]{-70.5}(-3.73,5.9){${q_2=q_1^2}$}
\psline[linecolor=black]{->}(0,-0.6)(0,6.3)
\psline[linecolor=black]{->}(-4,0)(4.2,0)
\rput(4.3,0.15){$q_1$}
\rput(-0.2,6){$q_2$}
\rput(-3,-0.3){$\kappa_{1}$}\rput(-1,-0.3){$\kappa_{2}$}\rput(0.4,-0.3){$\kappa_{3}$}\rput(1.8,-0.3){$\kappa_{4}$}
\rput(3.8,-0.3){$\kappa_{5}$}
\psdots(-3,0)(-1,0)(0.33,0)(1.8,0)(3.8,0)
\psdots(-3,3.4)(-1,0.4)(0.33,0.05)(1.8,1.25)(3.8,5.4)
\psline[linestyle=dashed](-3,0)(-3,3.4)\psline[linestyle=dashed](-1,0)(-1,0.4)\psline[linestyle=dashed](1.8,0)(1.8,1.25)
\psline[linestyle=dashed](3.8,0)(3.8,5.4)
\pspolygon(-3,3.4)(-1,0.4)(0.33,0.05)(1.8,1.25)(3.8,5.4)
\end{pspicture}
\end{center}
\caption{Polygon $\Po$ in the case $\No=5$}
\end{figure}

Now, taking (\ref{5.32}) into account we can prove the following result.

\begin{lemma}\label{lemma2}
Let the $\{N_a,N_b\}$-soliton potential $u(x)$ be such that         its $\tau$-function (\ref{tau}), (\ref{tauf1}) obeys condition that on any subset $\{n_1<\ldots<N_b\}\in\{1,\ldots,\No\}$ the ratio
\begin{equation}
\Biggl(\prod_{l=1}^{N_{b}}e^{K_{n_{l}}(x)}\Biggr)\Biggm/\tau(x)\label{c1}
\end{equation}
is bounded for all $x$ and has finite limits at space infinity.  Then $M_{\text{d}}(x,x';q)$ for all $q$ in the strip $\kappa_1\leq{q_1}\leq{\kappa_{\No}}$ and outside the polygon $\Po$, is a bounded function of all its arguments including values at infinities, provided that in (\ref{MDelta}) the upper sign is chosen for $q$ above the polygon $\Po$ and the bottom sign for $q$ below the polygon $\Po$.
\end{lemma}

\textsl{Proof.\/} Thanks to (\ref{5.32}) we consider only $q$ belonging to the strip $\kappa_1\leq{q_1}\leq{\kappa_{\No}}$. Let us denote $z_{mn}=x_{1}+(\kappa_{m}+\kappa_{n})x_{2}$. Then by means of the identity
\begin{align*}
\theta (q_{1}-\kappa_{m})-& \theta (q_{1}-\kappa_{n})=\sgn(z_{mn}-z_{mn}')\times\notag\\
& \times \left[\theta\bigl((q_{1}-\kappa_{m})(z_{mn}-z_{mn}')\bigr)-
\theta\bigl((q_{1}-\kappa_{n})(z_{mn}-z_{mn}')\bigr)\right],
\end{align*}
we rewrite (\ref{6}) in the form
\begin{align}
& M_{\text{d}}(x,x';q)=\dfrac{\mp\theta\bigl(\pm(x^{}_{2}-x'_{2})\bigr)}{((N_{b}-1)!)^{2}(N_{b}+1)\tau (x)\tau (x')}
\times  \notag \\
&\quad\times\sum_{\substack{m_{i},n_{i}=1,\ldots,\No\\
[\kappa_{m_{N_b}},\kappa_{n_{N_b}}]\supseteq[\kappa_{n},\kappa_{n+1}]}}\Do(\{m_{i}\})\Do(\{n_{i}\})
\sgn(z_{m_{N_{b}}n_{N_{b}}}-z'_{m_{N_{b}}n_{N_{b}}})\times  \notag \\
&\quad V(\{m_{i}\},n_{N_{b}})V(n_{1},n_{2},\ldots ,n_{N_{b}-1})\times  \notag \\
& \quad \times[\theta ((q_{1}-\kappa_{m_{N_{b}}})(z_{m_{N_{b}}n_{N_{b}}}^{}-z'_{m_{N_{b}}n_{N_{b}}}))-
\theta ((q_{1}-\kappa_{n_{N_{b}}})(z_{m_{N_{b}}n_{N_{b}}}^{}-z'_{m_{N_{b}}n_{N_{b}}}))]\times  \notag \\
& \quad \times \exp \Biggl(\,{\sum_{l=1}^{N_{b}}K_{m_{l}}(x)+K_{n_{N_{b}}}(x)+
\sum_{l=1}^{N_{b}-1}K_{n_{l}}(x')-q(x-x')}\,{\Biggr)}.\label{7}
\end{align}
We decompose this representation as a sum of two terms in correspondence to the two terms in the forth line,  we replace $q(x-x')$ in the exponential factors with the identity $q(x-x')=K_{m}(x)-K_{m}(x')+q_{mn}(x_{2}^{}-x'_{2})+(q_{1}-\kappa_{m})(z_{mn}^{}-z_{mn}')$, where $K_{m}(x)$ is defined in (\ref{Kn}) and $q_{mn}$ in (\ref{qmn}), and, finally, we choose $m=m_{N_b}$, $n=n_{N_b}$ in the first term and  viceversa in the second one. Thus we get
\begin{equation}
M_{\text{d}}(x,x';q)=M_{}^{(1)}(x,x';q)+M_{}^{(2)}(x,x';q),  \label{AP54}
\end{equation}
where in each substripe $\kappa_{n}\leq{q_1}\leq{\kappa_{n+1}}$, $n=1,\ldots,\No$
\begin{align}
&M_{}^{(1)}(x,x';q)=\dfrac{\mp\theta\bigl(\pm(x_{2}^{}-x'_{2})\bigr)}{((N_{b}-1)!)^{2}(N_{b}+1)}
\sum_{\substack{m_{i},n_{i}=1,\ldots,\No\\ [\kappa_{m_{N_b}},\kappa_{n_{N_b}}]\supseteq[\kappa_{n},\kappa_{n+1}]}}
\sgn(z_{m_{N_{b}}n_{N_{b}}}-z_{m_{N_{b}}n_{N_{b}}}')\times  \notag\\
&\quad\times e_{}^{-q_{m_{N_{b}}n_{N_{b}}}(x_{2}^{}-x_{2}')}V(\{m_{i}\},n_{N_{b}})V(n_{1},n_{2},\ldots,n_{N_{b}-1})
\times\notag \\
&\quad\times\theta\bigl((q_{1}-\kappa_{m_{N_{b}}})(z_{m_{N_{b}}n_{N_{b}}}^{}-z'_{m_{N_{b}}n_{N_{b}}})\bigr)
e_{}^{-(q_{1}-\kappa_{m_{N_{b}}})(z_{m_{N_{b}}n_{N_{b}}}^{}-z'_{m_{N_{b}}n_{N_{b}}})}\times  \notag \\
& \quad \times \dfrac{\Do(\{m_{i}\})\exp \Biggl(\,{\sum_{l=1}^{N_{b}-1}K_{m_{l}}(x)+K_{n_{N_{b}}}(x)}\,{\Biggr)}}
{\tau (x)}\times  \notag \\
& \quad \times \dfrac{\Do(\{n_{i}\})\exp \Biggl(\,{\sum_{l=1}^{N_{b}-1}K_{n_{l}}(x')+K_{m_{N_{b}}}(x')}\,{\Biggr)}}
{\tau (x')},  \label{8}
\end{align}
and
\begin{align}
&M_{}^{(2)}(x,x';q)=\dfrac{\pm\theta\bigl(\pm(x_{2}^{}-x_{2}')\bigr)}{((N_{b}-1)!)^{2}(N_{b}+1)}
\sum_{\substack{m_{i},n_{i}=1,\ldots,\No\\ [\kappa_{m_{N_b}},\kappa_{n_{N_b}}]\supseteq[\kappa_{n},\kappa_{n+1}]}}
\sgn(z_{m_{N_{b}}n_{N_{b}}}-z_{m_{N_{b}}n_{N_{b}}}')\times  \notag\\
&\quad\times e_{}^{-q_{m_{N_{b}}n_{N_{b}}}(x_{2}^{}-x_{2}')}V(\{m_{i}\},n_{N_{b}})V(n_{1},n_{2},\ldots,n_{N_{b}-1})
\times  \notag \\
& \quad \times \theta ((q_{1}-\kappa_{n_{N_{b}}})(z_{m_{N_{b}}n_{N_{b}}}^{}-z'_{m_{N_{b}}n_{N_{b}}}))
e_{}^{-(q_{1}-\kappa_{n_{N_{b}}})(z_{m_{N_{b}}n_{N_{b}}}^{}-z'_{m_{N_{b}}n_{N_{b}}})}\times  \notag \\
& \quad \times \dfrac{\Do(\{m_{i}\})\exp \Biggl(\,{\sum_{l=1}^{N_{b}}K_{m_{l}}(x)}\,{\Biggr)}}{\tau (x)}\times\notag\\
& \quad \times \dfrac{\Do(\{n_{i}\})\exp \Biggl(\,{\sum_{l=1}^{N_{b}}K_{n_{l}}(x')}\,{\Biggr)}}{\tau (x')}.\label{9}
\end{align}
We see that the dependence of these relations on space variables is due to the exponential factors and $\tau$-functions. On the other side, since the extended resolvent must belong to the space of tempered distributions, we have to show that these exponential factors cannot grow either thanks to the $\theta$-functions, or thanks to the $\tau$-functions in the denominators.

Let us consider the $x$-behavior of these two expressions in detail. First, recalling definition (\ref{qmn}) of $q_{mn}$, let us notice that, for $q$ belonging to the $n$-th substripe and $q_{n,n+1}\leq 0$, i.e., for $q$ in this substripe below the polygon $\Po$, or on its bottom border, all other $q_{m_{N_b},n_{N_b}}$ involved in the summation are nonpositive. As well, for  $q_{1,\No}\geq 0$, i.e., for $q$ in this stripe above the polygon $\Po$, or on its upper border, then all other $q_{m_{N_b},n_{N_b}}$ involved in the summation are nonnegative. Therefore, if one is choosing the signs in the r.h.s.\ of (\ref{8}) and (\ref{9}) as indicated in the Lemma, the exponents in the second lines are decaying or bounded when $x$ or $q$ tends to infinity. The exponents in the third lines of the r.h.s.\ of (\ref{8}) and (\ref{9}) are decaying or at least not growing thanks to the $\theta$-functions in the second lines. So we have to check the behavior with respect to $x$ and $x'$ of the last two lines of these equations. About (\ref{9}) the situation is trivial. The exponents in the forth and fifth lines have the same coefficient (minor of $\Do$) as in $\tau(x)$ and $\tau(x')$ in the denominator and, therefore, these terms are bounded when $x$ and $x'$ are growing. Situation with (\ref{8}) is more involved. Let us consider the term in the forth line. If the minor $\Do(\{m_{i}\})$ in the numerator is different from zero and $\Do(m_{1},\ldots,m_{N_{b}-1},n_{N_{b}})\neq 0$, then the same exponent as in the numerator is present in $\tau(x)$ in the denominator and the ratio is bounded. However, if $\Do(m_{1},\ldots,m_{N_{b}-1},n_{N_{b}})=0$ such exponent is not involved in $\tau(x)$ and in the direction where it is the leading one (if such direction exists) the ratio is growing at large space. The same is valid for the term in the fifth line of (\ref{8}). Thus boundedness of these ratios and then of the whole expression (\ref{8}) is guaranteed by the condition (\ref{c1}) of the lemma. $\blacksquare$

\begin{remark}
It is clear that condition (\ref{c1}) of this lemma is enough for the validity of its statement, but not necessary, and boundedness of $M_{\text{d}}(x,x';q)$ requires additional study. Nevertheless, it is already clear that the case of a Totally Positive (TP) matrix $\Do$ guarantees the implementation of condition (\ref{c1}). If instead of TP we impose on the matrix $\Do$ conditions (\ref{zn0}), then all leading exponents are involved in $\tau (x)$, as we mentioned in discussion of (\ref{3:232}) and (\ref{tauasympt}). Thus, (\ref{c1}) holds if $\tau(x)$ has no zeroes in the finite domain. To avoid this singularities it is enough to impose additionally that the matrix $\Do$ is TNN.
\end{remark}
\begin{remark}
Boundedness of $M(x,x';q)$ with respect to the variable $q$  when $q\in\Rs^{2}\setminus\Po$, on the boundaries of $\Po$, and in the limits at $q$-infinity follows from boundedness of $M_{\text{c}}(x,x';q)$ in (\ref{20:1})
and $M^{(1)}(x,x';q)$, $M^{(2)}(x,x';q) $ in (\ref{8}) and (\ref{9}). In the next section we consider the behavior of the function $M(x,x';q)$ with respect to $q$ in detail.
\end{remark}
\begin{remark}
Summarizing, we proved that under conditions of Lemma \ref{lemma2} the function $M(x,x';q)$ is a bounded function of its arguments and has finite asymptotic behavior. This means that this function belongs to the class of tempered distributions, i.e, it is the extended resolvent of the heat operator $\mathcal{L}(x,x';q)$ for $q$ outside the polygon $\mathcal{P}$.
\end{remark}

\section{Properties of the resolvent and Green's functions}
\subsection{Extended resolvent inside and outside parabola $q^{}_2=q_{1}^2$}

Local properties of the extended resolvent are easier to study in terms of the hat-kernel (\ref{Mhat}). Under the special reduction (\ref{GM}) this kernel is nothing but the total Green's function $\Go(x,x',\bk)$, where $\bk\in\Cs$ is the spectral parameter. In fact, since under this reduction $q_2-q_1^2=-\bk_{\Re}^{2}$ is nonpositive, and $q_{1\No}$ (see (\ref{qmn})) are less or equal to zero, it gives a mapping of the exterior of the parabola $q_2^{}=q_1^{2}$ region of the $q$ plane on the complex plane of $\bk$ (precisely, a one-to-two mapping since the reduction depends on $|\bk_{\Re}|$ and not on $\bk_{\Re}$). Taking (\ref{5.32}) into account we see that the part of the strip $\kappa_{1}\leq{q_1}\leq\kappa_{\No}$ outside of the parabola is below the polygon $\Po$, so we have to choose the bottom sign in (\ref{MDelta}). Thus, thanks to (\ref{M}),  (\ref{20}), and (\ref{Phat}) we get
\begin{align}
\Go(x,x',\bk) =&-\dfrac{\sgn(x_{2}-x_{2}')}{2\pi}\int d\alpha\,\theta\bigl((\alpha_{}^{2}-\bk_{\Re}^{2})(x_{2}-x_{2}')\bigr)
\Phi (x,\alpha +i\bk_{\Im})\Psi (x',\alpha +i\bk_{\Im})+\notag\\
&+i\theta(x_2'-x_2)\sum_{n=1}^{\No}\theta (\bk_{\Im}-\kappa_{n})\Phi (x,i\kappa_{n})\Psi_{n}(x'),\label{prop1}
\end{align}
that is the Green's function derived in \cite{KPIIGreen}.  Its boundedness property (see (\ref{Green}))  proved there follows now from boundedness of the resolvent at infinity.

The interior part of the parabola, as was already mentioned in discussion of (\ref{19}), is divided by the polygon $\Po$ in the part above the polygon ($q_{1,\No}\geq0$) and $\No-1$ lenses bounded by the parabola and its chords connecting points $q=(\kappa^{}_{n},\kappa_{n}^2)$ and $q=(\kappa^{}_{n+1},\kappa_{n+1}^2)$, $n=1,\ldots,\No-1$ (see Fig. \ref{figure}). All these lenses are below the polygon $\Po$. Taking into account that for $q_2^{}\geq{q_{1}^2}$ both $\h{M}_{\text{c}}(x,x';q)$ and $\h{M}_{\text{d}}(x,x';q)$ are independent of $q_2$ and in analogy to (\ref{GM}) we introduce
\begin{align}
\Go^{+}(x,x',\bk_{\Im})=\h{M}(x,x';q)\Bigm|_{q_{1}=\bk_{\Im}},\quad q^{}_{2}\geq{q_{1}^{2}},\quad
&q_2\geq(\kappa_{1}+\kappa_{\No})q_1-\kappa_{1}\kappa_{\No},\label{GM+}\\
\Go^{-}(x,x',\bk_{\Im})=\h{M}(x,x';q)\Bigm|_{q_{1}=\bk_{\Im}},\quad q^{}_{2}\geq{q_{1}^{2}},\quad
&q_2\leq(\kappa_{n}+\kappa_{n+1})q_1-\kappa_{n}\kappa_{n+1},\label{GM-}\\
&n=1,\ldots,\No.\nonumber
\end{align}
Again, thanks to (\ref{M}),  (\ref{20}), (\ref{Phat}) and (\ref{MDelta}) (once chosen the proper sign according to Lemma \ref{lemma2}) we derive that
\begin{align}
\Go^{+}(x,x',\bk_{\Im}) =&-\dfrac{\theta(x^{}_{2}-x_{2}')}{2\pi}\int d\alpha\,\Phi (x,\alpha +i\bk_{\Im})
\Psi (x',\alpha +i\bk_{\Im})-\notag\\
&-i\theta(x_2^{}-x'_2)\sum_{n=1}^{\No}\theta(\bk_{\Im}-\kappa_{n})\Phi (x,i\kappa_{n})\Psi_{n}(x'),\label{prop2}\\
\Go^{-}(x,x',\bk_{\Im}) =&-\dfrac{\theta(x^{}_{2}-x_{2}')}{2\pi}\int d\alpha\,\Phi (x,\alpha +i\bk_{\Im})
\Psi (x',\alpha +i\bk_{\Im})+\notag\\
&+i\theta(x'_2-x^{}_2)\sum_{n=1}^{\No}\theta(\bk_{\Im}-\kappa_{n})\Phi (x,i\kappa_{n})\Psi_{n}(x').\label{prop3}
\end{align}

Boundedness property for these Green's functions follows from the boundedness of the extended resolvent proved in Lemma \ref{lemma2}. Precisely, thanks to (\ref{Mhat}) and (\ref{GM+}), (\ref{GM-}) we have that
\begin{align}
e_{}^{\bk_{\Im}(x'_1-x^{}_1)+s(x'_2-x^{}_2)}&\Go^{+}(x,x',\bk_{\Im})\text{ is bounded for any real } s \text{ such that}\nonumber\\
& s\geq\bk_{\Im}^2+\max\{0,(\kappa_{\No}-\bk_{\Im})(\bk_{\Im}-\kappa_{1})\},\label{prop4}
\end{align}
and
\begin{align}
e_{}^{\bk_{\Im}(x'_1-x^{}_1)+s(x'_2-x^{}_2)}&\Go^{-}(x,x',\bk_{\Im})\text{ is bounded for any real } s \text{ such that}\nonumber \\
&\bk_{\Im}^2\leq{s}\leq\bk_{\Im}^2+(\kappa_{n+1}-\bk_{\Im})(\bk_{\Im}-\kappa_{n}),\quad n=1,\ldots,\No-1.\label{prop5}
\end{align}

\subsection{Discontinuities of the Green's functions}

In order to describe the discontinuities of the considered reductions of the resolvent, it is convenient to use for the resolvent representation (\ref{20:1}), so that, thanks to (\ref{Phi}), (\ref{Psi}), (\ref{Mhat}),  (\ref{M+MDelta}) and (\ref{Phat}),  (\ref{MDelta}), we can write
\begin{align}
\h{M}(x,x';q)&=\h{M}_{0}(x,x';q)-\nonumber\\
&-\sum_{n=1}^{\No}\Phi(x,i\kappa_{n})\Psi_{n}(x')\Biggl\{
\dfrac{\sgn(x_{2}-x_{2}')}{2\pi}\int d\alpha\dfrac{\theta \bigl((q_{2}+\alpha^{2}-q_{1}^{2})(x^{}_{2}-x_{2}')\bigr)}
{\alpha+i(q_1-\kappa_{n})}\times\nonumber\\
&\times e_{}^{(q_1-\kappa_{n}-i\alpha)(x^{}_1-x'_1)-[(\alpha+iq_{1})^{2}+\kappa_{n}^{2}](x_{2}-x_{2}')}\pm
i\theta \bigl(\pm(x_{2}-x_{2}')\bigr)\theta (q_{1}-\kappa_{n})\Biggr\},\label{prop6}
\end{align}
where by (\ref{Mhat}) and (\ref{20:2})
\begin{align}
\h{M}_{0}(x,x';q)=-\dfrac{\sgn(x_{2}-x_{2}')}{2\pi}\int d\alpha\,&\theta\bigl((q_{2}+\alpha^{2}-q_{1}^{2})(x^{}_{2}-x_{2}')\bigr)\times\nonumber\\
&\times e_{}^{(q_1-i\alpha)(x^{}_1-x'_1)+(q_1-i\alpha)^2(x^{}_2-x'_2)},\label{prop7}
\end{align}
and where the sign in the r.h.s.\ of (\ref{prop6}) must be chosen in agreement with Lemma \ref{lemma2}. It is easy to see, that outside the parabola $q^{}_{2}=q_{1}^{2}$ the kernel $\h{M}(x,x';q)$ is a continuous function of $q$, when $q\neq(\kappa^{}_{n},\kappa_{n}^2)$, $n=1,\ldots,\No$.

From (\ref{prop6}), by using the reduction (\ref{GM}), we get for the total Green's function given in (\ref{prop1}) the following representation
\begin{align}
\Go(x,x',\bk)&=-\dfrac{\sgn(x_{2}-x_{2}')}{2\pi}\int\limits_{\bk'_{\Im}=\bk^{}_{\Im}}\!\!\!\! d\bk'_{\Re}\,
\theta\bigl(({\bk'_{\Re}}^{2}-{\bk^{}_{\Re}}^{2})(x^{}_{2}-x_{2}')\bigr)e_{}^{-i\bk'(x^{}_1-x'_1)-{\bk'}^2(x^{}_2-x'_2)}    -\nonumber\\
&-\sum_{n=1}^{\No}\Phi(x,i\kappa_{n})\Psi_{n}(x')\Biggl\{
\dfrac{\sgn(x_{2}-x_{2}')}{2\pi}\int\limits_{\bk'_{\Im}=\bk^{}_{\Im}}\!\!\!\! d\bk'_{\Re}\,\dfrac{\theta \bigl(({\bk'_{\Re}}^{2}-{\bk^{}_{\Re}}^{2})(x^{}_{2}-x_{2}')\bigr)}
{\bk'-i\kappa_{n})}\times\nonumber\\
&\times e_{}^{-(\kappa_{n}+i\bk')(x^{}_1-x'_1)-(\kappa_{n}^{2}+{\bk'}^2_{})(x_{2}-x_{2}')}-
\dfrac{i}{2}\theta(x'_{2}-x^{}_{2})\sgn (\bk_{\Im}-\kappa_{n})\Biggr\},\label{prop61}
\end{align}
where  for $p(x.x';q_1)$ we used expression in (\ref{symPhat}).

Thanks to the above discussion, it is a continuous function of $\bk\in\Cs$ for all $\bk\neq{i\kappa_{n}}$. In order to study  the behavior of the Green's functions nearby these points, we consider first the Green's function $G^{+}$, as given in (\ref{prop2}). Thanks to (\ref{prop6}) we readily get
\begin{align}
&\Go^{+}(x,x',\bk_{\Im})=-\dfrac{\theta(x_{2}-x_{2}')}{2\pi}\int\limits_{\bk'_{\Im}=\bk^{}_{\Im}}\!\!\!\! d\bk'_{\Re}\,
e_{}^{-i\bk'(x^{}_1-x'_1)-{\bk'}^2(x^{}_2-x'_2)}-\sum_{n=1}^{\No}\Phi(x,i\kappa_{n})\Psi_{n}(x')\times\nonumber\\
&\qquad\times\theta(x_{2}-x_{2}')\Biggl\{\dfrac{1}{2\pi}\int\limits_{\bk'_{\Im}=\bk^{}_{\Im}}\!\!\!\! d\bk'_{\Re}\,
\dfrac{e_{}^{-(\kappa_{n}+i\bk')(x^{}_1-x'_1)-(\kappa_{n}^{2}+{\bk'}^2_{})(x^{}_2-x'_2)}}{\bk'-i\kappa_{n}}+
\dfrac{i}{2}\sgn (\bk_{\Im}-\kappa_{n})\Biggr\}.\label{prop8}
\end{align}

It is easy to see that this Green's function is continuous with respect to the variable $\bk_{\Im}$, and, moreover,  $\Go^{+}(x,x',\bk_{\Im})$ is independent of $\bk_{\Im}$. Indeed, differentiating by $\bk_{\Im}$, we use that exponents are analytic with respect to the variable $\bk'=\bk'_{\Re}+i\bk_{\Im}$ and rapidly decay when $\bk'_{\Re}$ tends to infinity, and also that $\overline{\partial}_{\bk'}(\bk'-i\kappa_{n})^{-1}=\pi\delta(\bk'_{\Re})\delta(\bk'_{\Im}-\kappa_{n})$. Thus the derivatives of the first term and every term in the parenthesis are equal to zero, so that we can write that
\begin{align}
\Go^{+}(x,x')&=-\dfrac{\theta(x_{2}-x_{2}')}{2\pi}\Biggl\{\int d\alpha\,
e_{}^{-i\alpha(x^{}_1-x'_1)-\alpha^2(x^{}_2-x'_2)}+\nonumber\\
&+\sum_{n=1}^{\No}\Phi(x,i\kappa_{n})\Psi_{n}(x')\int \dfrac{d\alpha}{\alpha}\,
e_{}^{-i\alpha(x^{}_1-x'_1+2\kappa_{n}(x^{}_2-x'_2))-\alpha^{2}(x^{}_2-x'_2)}\Biggr\},\label{prop81}
\end{align}
where integrals are understood in the sense of the principal value and that also can be calculated explicitly in terms of the hypergeometric functions.

On the other side, directly by (\ref{prop2}) and (\ref{prop3}) we have that
\begin{equation}
\Go^{-}(x,x',\bk_{\Im})=\Go^{+}(x,x')+
\dfrac{i}{2}\sum_{n=1}^{\No}\sgn(\bk_{\Im}-\kappa_{n})\Phi(x,i\kappa_{n})\Psi_{n}(x').\label{prop9}
\end{equation}
So $\Go^{-}(x,x',\bk_{\Im})$ is discontinuous at all $\bk_{\Im}=\kappa_{n}$, and, precisely,
\begin{equation}
\Go^{-}(x,x',\kappa_{n}+0)-\Go^{-}(x,x',\kappa_{n}-0)=i\Phi(x,i\kappa_{n})\Psi_{n}(x'),
\quad n=1,\ldots,\No.\label{prop10}
\end{equation}

Thus, considering the difference of the expressions given in (\ref{prop61}) and (\ref{prop8}) we derive that the total Green's function can be written in the form
\begin{equation}
\Go(x,x',\bk)=\Go_{\reg}(x,x',\bk)+\Go_{\Delta}(x,x',\bk), \label{prop11}
\end{equation}
where
\begin{align}
\Go_{\reg}&(x,x',\bk)=\Go^{+}(x,x')+\dfrac{1}{2\pi}
\int\limits_{\substack{-|\bk_{\Re}|\\ \bk'_{\Im}=\bk^{}_{\Im}}}^{|\bk_{\Re}|}\!\!\!\! d\bk'_{\Re}\,
e_{}^{-i\bk'(x^{}_1-x'_1)-{\bk'}^2(x^{}_2-x'_2)}+\nonumber\\
&+\dfrac{1}{2\pi}\sum_{n=1}^{\No}\Phi(x,i\kappa_{n})\Psi_{n}(x')
\int\limits_{\substack{-|\bk_{\Re}|\\ \bk'_{\Im}=\bk^{}_{\Im}}}^{|\bk_{\Re}|}\!\!\!\! d\bk'_{\Re}\,
\dfrac{e_{}^{-(\kappa_{n}+i\bk')(x^{}_1-x'_1)-(\kappa_{n}^{2}+{\bk'}^2_{})(x^{}_2-x'_2)}-1}{\bk'-i\kappa_{n}},
\label{prop12}
\end{align}
the last integral term being regularized by a subtraction of 1 in the numerator that compensates the zero in the denominator at $\bk'=i\kappa_{n}$. Correspondingly,
\begin{equation}
\Go_{\Delta}(x,x',\bk)=\dfrac{1}{i\pi}\sum_{n=1}^{\No}\Phi(x,i\kappa_{n})\Psi_{n}(x')
\cot^{-1}\dfrac{\bk_{\Im}-\kappa_{n}}{|\bk_{\Re}|},\label{prop13}
\end{equation}
where we omitted a term proportional to the l.h.s.\ of (\ref{sumPhiPsi}). This relation is explicitly discontinuous at every $\bk=i\kappa_{n}$ and only at these points. Precisely,
\begin{align}
\Go_{\Delta}(x,x',\bk)&=\dfrac{1}{i\pi}\Phi(x,i\kappa_{n})\Psi_{n}(x')
\cot^{-1}\dfrac{\bk_{\Im}-\kappa_{n}}{|\bk_{\Re}|}-\nonumber\\
&-i\sum_{m=n+1}^{\No}\Phi(x,i\kappa_{m})\Psi_{m}(x')+o(1),\quad\bk\sim i\kappa_{n},\quad n=1,\ldots,\No.
\label{prop14}
\end{align}
Let us also mention that thanks to (\ref{prop11})--(\ref{prop13})
\begin{equation*}
\Go(x,x',\bk)\bigm|_{\bk_{\Re}=0}=\Go^{+}(x,x')-\dfrac{i}{2}
\sum_{n=1}^{\No}\sgn(\kappa_{n}-\bk_{\Im})\Phi(x,i\kappa_{n})\Psi_{n}(x'),
\end{equation*}
where it is assumed that $\bk_{\Im}\neq\kappa_{n}$, $n=1,\ldots,\No$. Thanks to (\ref{sumPhiPsi}) and (\ref{prop9}) this gives
\begin{equation}
\Go(x,x',\bk)\bigm|_{\bk_{\Re}=0}=\Go^{-}(x,x',\bk_{\Im}).\label{prop15}
\end{equation}
On the other side, in vicinity of the points $i\kappa_{n}$ we have
\begin{align}
\Go(x,x',\bk)&=\Go^{+}(x,x')+\dfrac{1}{i\pi}\Phi(x,i\kappa_{n})\Psi_{n}(x')\cot^{-1}
\dfrac{\bk_{\Im}-\kappa_{n}}{|\bk_{\Re}|}-\nonumber\\
&-i\sum_{m=n+1}^{\No}\Phi(x,i\kappa_{m})\Psi_{m}(x')+o(1),\quad\bk\sim i\kappa_{n},\quad n=1,\ldots,\No.
\label{prop16}
\end{align}
In a forthcoming publication we will show that these properties of the Green's functions enable the solution of the problem formulated in the Introduction, i.e., to develop the IST for the case of perturbed multisoliton potentials (\ref{tu}) of the heat operator (\ref{heatop}).

\section*{Acknowledgments}

This work is supported in part by the grant RFBR \# 11-01-00440, Scientific Schools 4612.2012.1, by the Program of RAS ``Mathematical Methods of the Nonlinear Dynamics,'' by INFN, by MIUR (grant PRIN 2008 ``Geometrical methods in the theory of
nonlinear integrable systems''), and by Consortium E.I.N.S.T.E.IN.

\end{document}